\documentstyle[a4,12pt, axodraw]{article}
\def\barr{\overline}

\makeatletter
\def\vereq#1#2{\lower3pt\vbox{\baselineskip1.5pt \lineskip1.5pt
\ialign{$\m@th#1\hfill##\hfil$\crcr#2\crcr\sim\crcr}}}
\makeatother

\begin{document}

\begin{titlepage}

\begin{flushright}
UCB-PTH-01/31 \\
LBNL-48728 \\
\end{flushright}

\vskip 1.5cm

\begin{center}
{\Large \bf SO(10) Unified Theories in Six Dimensions}

\vskip 1.0cm

{\large
Lawrence Hall, Yasunori Nomura, Takemichi Okui and David Smith
}

\vskip 0.5cm

{\it Department of Physics, 
 \\ and \\
 Theoretical Physics Group, Lawrence Berkeley National Laboratory,\\
 University of California, Berkeley, CA 94720}

\vskip 1.0cm

\abstract{We construct supersymmetric models of $SO(10)$ unification in
 which the gauge symmetry is broken by orbifold compactification.  We
 find that using boundary conditions to break the gauge symmetry down to 
 $SU(3)_C \otimes SU(2)_L \otimes U(1)_Y \otimes U(1)_X$ without leaving 
 unwanted massless states requires at least two extra dimensions,
 motivating us to work with 6D orbifolds.  $SO(10)$ is broken by two
 operations, each of which induces gauge-breaking to either the
 Georgi-Glashow, Pati-Salam, or flipped $SU(5) \otimes U(1)$ subgroups;
 assigning different unbroken subgroups to the two operations leaves
 only the standard model gauge group and $U(1)_X$ unbroken.  The models
 we build employ extra-dimensional mechanisms for naturally realizing
 doublet-triplet splitting, suppressing proton decay, and avoiding
 unwanted grand-unified fermion mass relations.  We find some tension
 between being free of anomalies of the 6D bulk, accommodating a simple
 mechanism for generating right-handed neutrino masses, and preserving
 the precise prediction of the weak mixing angle.}

\end{center}
\end{titlepage}

\section{Introduction}

The successful prediction of the weak mixing angle in the minimal
supersymmetric standard model (MSSM) is a compelling hint for new
physics.  The most direct interpretation of this hint is that of low
energy supersymmetry and an energy desert, with no additional physics,
extending between the TeV scale and the unification scale 
$M_U \sim 10^{16}$ GeV \cite{Dimopoulos:1981yj}.

How can nature be described above $M_U$?  One possibility is that
it is described by a grand unified theory (GUT) \cite{Georgi:1974sy, 
Dimopoulos:1981zb}.  Grand unification offers an elegant explanation of
the quantum numbers of the standard model quarks and leptons, but raises
other new questions.  These include the details of the gauge symmetry
breaking, the origin of doublet-triplet splitting, and the reason for
non-observation of proton decay.

There has been much recent interest in addressing these issues
in the context of grand unified theories with extra spacetime
dimensions \cite{Kawamura:2001ev, Altarelli:2001qj, Hall:2001pg,
Hebecker:2001wq, Barbieri:2001yz, Kawamura:2001ir, Nomura:2001mf, 
Barbieri:2001dm, Hebecker:2001jb, Hall:2001tn, Weiner:2001pv, 
Hall:2001zb}.  These recent models apply ideas that first appeared in
string-motivated work \cite{Candelas:1985en}: the gauge symmetry is
broken by identifications imposed on the gauge fields under the
spacetime symmetries of an orbifold, and doublet-triplet splitting
occurs because the orbifold compactification projects out the zero modes 
of the colored components of the Higgs multiplets.  
In these models, however, there is  
a moderately large energy interval where the physics is described by a
higher-dimensional grand unified field theory,
and this mild hierarchy between the cutoff and compactification
scales is crucial for guaranteeing
the smallness of threshold corrections to 
$\sin^2 \theta_w$ \cite{Hall:2001pg}.
The absence of proton decay induced by dimension five operators can also
be given an intrinsically extra dimensional explanation involving the
form of the mass matrix for the Higgsino Kaluza-Klein (KK) modes
\cite{Hall:2001pg}.

These ideas have been used to build complete and realistic 5D models of
supersymmetric $SU(5)$ unification on an $S^1/Z_2$ orbifold 
\cite{Kawamura:2001ev, Altarelli:2001qj, Hall:2001pg, Hebecker:2001wq,
Barbieri:2001yz}.  The purpose of this paper is to explore whether
similar ideas can be used to build simple models based on $SO(10)$
gauge symmetry.  One motivation for considering $SO(10)$ carries over
from the 4D case: $SO(10)$ allows an entire generation of quarks and
leptons to be unified in an irreducible spinor representation.  This
representation includes a right-handed neutrino, so that $SO(10)$ also
provides a natural framework within which the see-saw mechanism
\cite{Seesaw} can be realized.  In the context of extra dimensional
models, we will find that working with $SO(10)$ also illustrates how
interesting group-theoretic structure can arise on orbifolds.  For
instance, the identifications we impose on the gauge fields under
spacetime symmetries to break $SO(10) \rightarrow SU(3)_C \otimes
SU(2)_L \otimes U(1)_Y \otimes U(1)_X$ naturally lead to fixed points in
which only the Pati-Salam $SU(4)_C \otimes SU(2)_L \otimes SU(2)_R$
\cite{Pati:1974yy}, Georgi-Glashow $SU(5) \otimes U(1)_X$
\cite{Georgi:1974sy}, or flipped $SU(5)' \otimes U(1)'_X$ \cite{Barr:1982qv}
subgroups of $SO(10)$ are preserved.  What is the minimum
number of extra dimensions required to break the gauge symmetry through
orbifold compactification?  In the $SU(5)$ case a single extra dimension
is sufficient.  We will find that the larger $SO(10)$ gauge group
requires at least two extra dimensions for the orbifold compactification
to break the unified symmetry without leaving extra massless states
coming from the higher-dimensional supersymmetric vector
multiplet. Thus the models we construct will be six dimensional.

The outline of the paper is as follows.  In section \ref{sec:group}, we
consider the group theoretic structure of $SO(10)$ gauge symmetry
breaking on a torus. The basic ideas discussed in that section are then
applied in the rest of the paper to construct three different orbifold
models.  The first, presented in section \ref{section:z2}, is a theory
with $N=1$ supersymmetry on a $T^2/Z_2$ orbifold.  We will find that
this orbifold provides a natural setting for doublet-triplet splitting
and for extra-dimensional mechanisms for relaxing unwanted grand unified 
fermion mass relations.  It also accommodates simple ways of breaking 
the $U(1)_X$ gauge symmetry left after orbifolding and communicating 
this breaking to give right-handed neutrino masses.  
The irreducible gauge anomalies for this theory are easily canceled
by choosing appropriate bulk matter content,
but the Green-Schwarz mechanism \cite{Green:1984sg}
is required to cancel the rest of the
anomaly, leading to axion-like states in the low-energy theory. This
motivates us to construct completely anomaly-free theories with 6D $N=2$
supersymmetry in sections \ref{section:t2z6} and \ref{section:z2sq}.
The model of section \ref{section:t2z6} is constructed on $T^2/Z_6$,
with only the 6D $N=2$ vector multiplet allowed in the bulk.  In this
model there are no colored Higgs multiplets: matter and Higgs are
localized to a fixed point that preserves only the Pati-Salam subgroup
of $SO(10)$, and the Higgs doublets are contained in the 
$({\bf 1}, {\bf 2} ,{\bf 2})$ representation.  The breaking of $U(1)_X$
is straightforward but communicating it to standard model fields is not,
and we will find that it is difficult to obtain right-handed neutrino
masses (and to avoid $SO(10)$ mass relations) in this model without
facing a vacuum alignment problem.  In section \ref{section:z2sq} we
attempt to improve this situation by working with a 
$T^2/(Z_2 \times Z_2')$ orbifold, which has 5D ``fixed lines'' on which
matter may propagate, without introducing 6D anomalies.  The existence
of these lines makes communicating $U(1)_X$ breaking and correcting
fermion mass relations much easier.  However, this gain is likely at the 
expense of the precise prediction of the weak mixing angle.
This issue is discussed in section \ref{sec:gcu}.  Our conclusions
appear in section \ref{sec:concl}.

During preparation of this manuscript, we received
Ref.~\cite{Asaka:2001eh}, which also considers $SO(10)$ breaking by
orbifold compactification in six dimensions.

\section{SO(10) Gauge Symmetry Breaking on a Torus}
\label{sec:group}

In this section we consider the $SO(10)$ breaking by orbifold
compactifications.  We begin by considering the case of a single extra
dimension.  The most general spacetime symmetries that can be used to
compactify a single extra dimension may be taken to be a reflection
${\cal Z}$ and a translation ${\cal T}$ \cite{Barbieri:2001dm}.  Fields
propagating in the extra dimension may transform nontrivially under
${\cal Z}$ and/or ${\cal T}$, as long as the bulk action is invariant
under these operations and the transformations under ${\cal Z}$ and
${\cal T}$ are consistent: ${\cal T}{\cal Z}$ and ${\cal Z}{\cal
T}^{-1}$ must act on fields in the same way because they induce the same
motion in spacetime.

We first ask whether we can build a 5D $N=1$ supersymmetric model, in
which $SO(10)$ is broken by these transformations to $SU(3)_C \otimes
SU(2)_L \otimes U(1)_Y \otimes U(1)_X$ (3-2-1-1).  Such breaking
requires both ${\cal Z}$ and ${\cal T}$ to have non-trivial gauge
properties; for example, ${\cal Z}$ and ${\cal T}$ may be chosen to
preserve $SU(4)_C \otimes SU(2)_L \otimes SU(2)_R$ (4-2-2) and 
$SU(5) \otimes U(1)_X$ (5-1) subgroups of $SO(10)$, respectively.
However, this results in the chiral adjoint of the 5D vector multiplet
containing extra massless fields other than the states in the MSSM, so
that the gauge coupling unification is spoiled.

In 6D this problem is immediately avoided, as there are now two
translations, ${\cal T}_1$ and ${\cal T}_2$, and they can be used to
break $SO(10)$ to $SU(3)_C \otimes SU(2)_L \otimes U(1)_Y \otimes
U(1)_X$.  The spacetime orbifold depends not only on the torus defined
by ${\cal T}_1$ and ${\cal T}_2$, but also on the non-freely acting
symmetries used to identify parts of the torus.  Here we assume that
these non-freely acting orbifold symmetries preserve $SO(10)$, and
hence for the purpose of describing the gauge symmetry breaking in this
section we need not discuss them.  In the next three sections different
orbifolds are constructed, and in each case the orbifolding symmetries
are used to ensure that there are no unwanted zero-mode states from the
6D vector supermultiplet.\footnote{
In the model of section \ref{section:t2z6}, the orbifold symmetry 
breaks $SO(10)$ while accomplishing this task, but the discussion here 
in terms of torus translations will be sufficient for illustrating 
the basic ideas we use for breaking $SO(10)$ down to 
$SU(3)_C \otimes SU(2)_L \otimes U(1)_Y \otimes U(1)_X$.}

The generators $T^a$ of SO(10) are imaginary and antisymmetric 
$10 \times 10$ matrices.  We will find it convenient to write these
generators as tensor products of $2 \times 2$ and $5 \times 5$ matrices, 
giving $\sigma_0 \otimes A_5$, $\sigma_1 \otimes A_5$, $\sigma_2
\otimes S_5$ and $\sigma_3 \otimes A_5$ as a complete set.  Here
$\sigma_0$ is the $2 \times 2$ unit matrix and $\sigma_{1,2,3}$ are the
Pauli spin matrices; $S_5$ and $A_5$ are $5 \times 5$ matrices that are
real and symmetric, and imaginary and antisymmetric, respectively.  The
$\sigma_0 \otimes A_5$ and $\sigma_2 \otimes S_5$ generators form an
$SU(5) \otimes U(1)_X$ subgroup of $SO(10)$, with $U(1)_X$ given by
$\sigma_2 \otimes I_5$.  We choose our basis so that the standard model
gauge group is contained in this $SU(5)$, with $SU(3)_C$ contained in
$\sigma_0 \otimes A_3$ and $\sigma_2 \otimes S_3$ and $SU(2)_L$
contained in $\sigma_0 \otimes A_2$ and $\sigma_2 \otimes S_2$, where
$A_3$ and $S_3$ have indices 1,2,3 and $A_2$ and $S_2$ have indices
4,5.  The generators of this $SU(5) \otimes U(1)_X$ subgroup can be
grouped as 
\begin{equation}
\hspace{-1in}
  SU(5) \otimes U(1)_X:
\begin{array}{ccc}
  \sigma_0 \otimes A_3 & \sigma_0 \otimes A_2 & \sigma_0 \otimes A_X \\
  \sigma_2 \otimes S_3 & \sigma_2 \otimes S_2 & \sigma_2 \otimes S_X.
\end{array}
\label{eq:51gen}
\end{equation} 
Here $A_X$ and $S_X$ denote the off diagonal pieces left over from $A_5$
and $S_5$. A different $SU(5) \otimes U(1)$ subgroup is formed by
replacing $\sigma_0 \otimes A_X$ and $\sigma_2 \otimes S_X$ with
$\sigma_1 \otimes A_X$ and $\sigma_3 \otimes A_X$: 
\begin{equation}
\hspace{-1in}
  SU(5)'\otimes U(1)_X':
\begin{array}{ccc}
  \sigma_0 \otimes A_3 & \sigma_0 \otimes A_2 & \sigma_1 \otimes A_X \\
  \sigma_2 \otimes S_3 & \sigma_2 \otimes S_2 & \sigma_3 \otimes A_X.
\end{array}
\label{eq:flippedgen}
\end{equation} 
This $SU(5)'$ is known in the literature as flipped $SU(5)$
\cite{Barr:1982qv}. It contains $SU(3)_C$ and $SU(2)_L$ but not
$U(1)_Y$.  Finally, it will be useful to list the generators that form
the $SU(4)_C \otimes SU(2)_L \otimes SU(2)_R$ subgroup of $SO(10)$:
\begin{equation}
\hspace{-1in}
  SU(4)_C \otimes SU(2)_L \otimes SU(2)_R:
\begin{array}{cc}
  (\sigma_0,\sigma_1,\sigma_3) \otimes A_3 & 
     (\sigma_0,\sigma_1,\sigma_3) \otimes A_2  \\
  \sigma_2 \otimes S_3 & \sigma_2 \otimes S_2. 
\end{array}
\label{eq:422gen}
\end{equation} 

The torus $T^2$ has translation symmetries defined by two vectors
$e_1$ and $e_2$ in the complex plane $z=x_5+ix_6$.  The translation
symmetries of the torus identify two points in the complex plane,
$z_1$ and $z_2$, if $z_1=z_2+m e_1+n e_2$ for integers $m$ and $n$.
Under the translation $z \rightarrow z+e_i$, the identifications imposed
on the vector supermultiplet, which contains the gauge fields, are
\begin{eqnarray}
  V(z+e_i) = T_i V(z) T_i^{-1}.
\end{eqnarray}
In this paper we employ three possible forms for the $T_i$ matrices.
They are
\begin{eqnarray}
  T_{51}   &\equiv& \sigma_2 \otimes I_5, \\
  T_{5'1'} &\equiv& \sigma_2 \otimes {\rm diag}(1,1,1,-1,-1), \\
  T_{422}  &\equiv& \sigma_0 \otimes {\rm diag}(1,1,1,-1,-1).
\end{eqnarray}
Consider, for instance, the case $(T_1,T_2)=(T_{51},T_{5'1'})$.
With this choice for $T_1$, we have  
\begin{equation}
\begin{array}{ccc}
  T_1 (\sigma_0 \otimes A_5) T_1^{-1} = 
       \sigma_0 \otimes A_5, 
& &
  T_1 (\sigma_1 \otimes A_5) T_1^{-1} = 
      -\sigma_1 \otimes A_5, 
\\ 
  T_1 (\sigma_2 \otimes S_5) T_1^{-1} = 
       \sigma_2 \otimes S_5, 
& &
  T_1 (\sigma_3 \otimes A_5) T_1^{-1} = 
      -\sigma_3 \otimes A_5.
\end{array}
\end{equation}
Thus, only $SU(5) \otimes U(1)_X$ gauge fields are potentially massless
once this transformation under the $e_1$ translation has been imposed.
On the other hand, the generators that commute with $T_2$ are
\begin{equation}
\begin{array}{ccc}
  \sigma_0 \otimes A_3 & \sigma_0 \otimes A_2 & \sigma_1 \otimes A_X \\
  \sigma_2 \otimes S_3 & \sigma_2 \otimes S_2 & \sigma_3 \otimes A_X,
\end{array}
\end{equation}
while all the other generators anticommute with $T_2$. 
Comparing with Eq.~(\ref{eq:flippedgen}), we see that the gauge fields
with even parity under the $e_2$ translation belong to $SU(5)' \otimes
U(1)'_X$ ($5'$-$1'$).  Combining with the result from the $e_1$
translation, we find that the only generators that are invariant under
both translations are those of $SU(3)_C \otimes SU(2)_L \otimes U(1)_Y
\otimes U(1)_X$. Therefore, only gauge fields from this subgroup will
have massless zero modes, as desired.  It is easily checked that taking 
$(T_1,T_2)=(T_{51},T_{422})$ or $(T_1,T_2)=(T_{5'1'},T_{422})$ leads
to the same unbroken gauge group.

\begin{figure}
\begin{center}
\begin{picture}(400,200)(0,0)
\Boxc(200,125)(200,150)
\Boxc(200,112)(400,124)
\Line(100,50)(70,0)
\Line(300,50)(270,0)
\Line(70,0)(270,0)
\Line(300,50)(330,10)
\Line(330,10)(276,10)
\DashLine(100,50)(130,10){3}
\DashLine(130,10)(276,10){3}
\Text(305,14)[b]{\small{$SU(2)_R$}}
\Text(100,4)[b]{\small{$SU(2)_L$}}
\Text(200,112)[c]{$U(3)_C$}
\Text(200,188)[c]{$SU(4)_C/U(3)_C$}
\Text(50,112)[c]{$5'/$(3-2-1)}
\Text(350,112)[c]{$5/$(3-2-1)}
\end{picture}
\caption{A convenient grouping of the $SO(10)$ generators.  The
parities of the corresponding gauge bosons under torus translations
are given in Eq.~(\ref{eq:gaugemodes}).}
\label{fig:generators}
\end{center}
\end{figure}
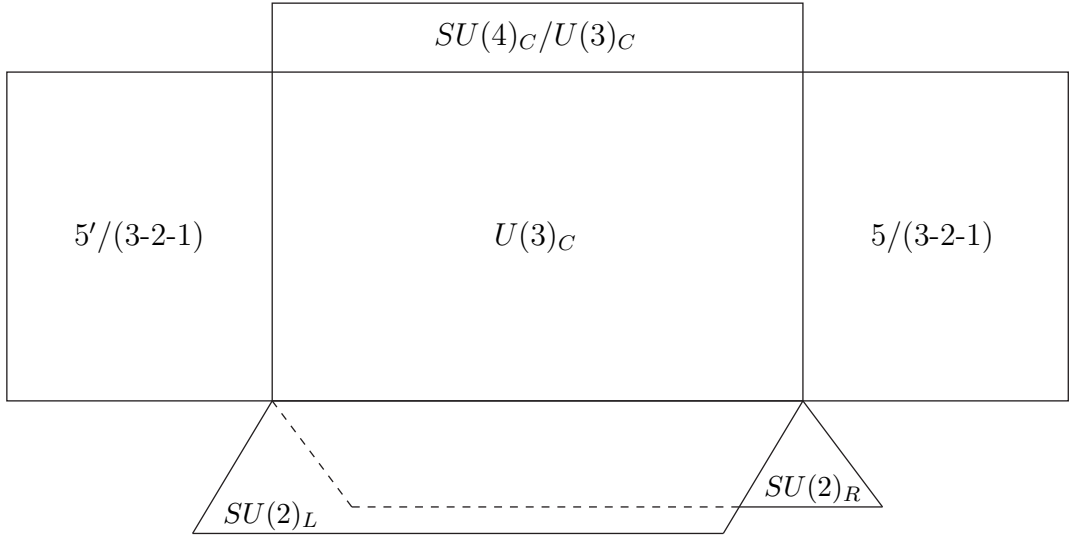

We finally summarize the group theoretic structure of $SO(10)$.
The 45 generators of $SO(10)$ can conveniently be assembled into seven
groups, as shown in Fig.~\ref{fig:generators}. 
The generators and their $(T_{51}, T_{5'1'})$ parities are given by 
\begin{equation}
\begin{array}{lll}
  U(3)_C  & \sigma_2 \otimes S_3;\; \sigma_0 \otimes A_3                   &
      (+,+) \\
  SU(2)_L & \sigma_2 \otimes \sigma^{\prime}_{1,3};\; \sigma_0 \otimes A_2 &
      (+,+) \\
  T_{3R}  & \sigma_2 \otimes \sigma^{\prime}_0                             &
      (+,+) \\
  T_R^\pm & \sigma_{1,3} \otimes A_2                                       &
      (-,-) \\
  SU(4)_C/U(3)_C & \sigma_{1,3} \otimes A_3                                &
      (-,-) \\
  SU(5)/(SU(3)_C \otimes SU(2)_L \otimes U(1)_Y)  & 
      \sigma_2 \otimes S_X;\; \sigma_0 \otimes A_X & (+,-) \\
  SU(5)'/(SU(3)_C \otimes SU(2)_L \otimes U(1)_Y) & 
      \sigma_{1,3} \otimes A_X                     & (-,+),
\end{array}
\label{eq:gaugemodes}
\end{equation}
where $\sigma^{\prime}_{0,1,3}$ are the components of $S_2$: 
$S_2 = \{ \sigma^{\prime}_0, \sigma^{\prime}_1, \sigma^{\prime}_3 \}$.
Here $U(3)_C$ contains $SU(3)_C$, and $SU(4)_C$ is the Pati-Salam
group. The generators of $U(1)_Y$ and $U(1)_X$ are linear combinations
of $T_{3R}$ and $U(3)_C/SU(3)_C$.

This figure represents well the symmetries among the $SO(10)$
generators.  There are symmetries which interchange $SU(2)_L$ and
$SU(2)_R$, and $SU(5)$ and $SU(5)'$.  It is also useful in identifying
the unbroken generators in patterns of $SO(10)$ breakings; 4-2-2 type
breaking corresponds to breaking generators in both left and right
wings, and 5-1 ($5'$-$1'$) type breaking to taking the body, the right
(left) wing, front leg, and $T_{3R}$ for unbroken generators.
Therefore, it is easily seen that the combination of any two of 4-2-2,
5-1, and  $5'$-$1'$ type breakings leads to $SU(3)_C \otimes SU(2)_L
\otimes U(1)_Y \otimes U(1)_X$ as the unbroken subgroup.

\section{A Model on $T^2/Z_2$}
\label{section:z2}

The first model we consider is a 6D $N=1$ supersymmetric model with the
extra dimensions compactified on a $T^2/Z_2$ orbifold.  In addition to
the identifications under torus translations, two points $z_1$ and $z_2$
are identified if they are mapped into each other under a $\pi$ rotation
in the $x_5$-$x_6$ plane, {\em i.e.} if $z_1=-z_2$.

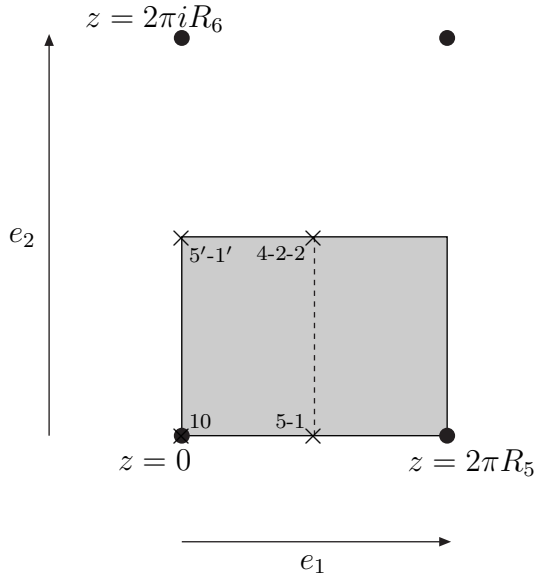
\begin{figure}
\begin{center}
\begin{picture}(180,220)(0,0)
\LongArrow(60,10)(160,10)
\GBox(60,50)(160,125){0.8}
\LongArrow(10,50)(10,200)
\DashLine(110,50)(110,125){2}
\Text(63,53)[bl]{\bf {\scriptsize 10}}
\Text(63,122)[tl]{\bf {\scriptsize 5$'$-1$'$}}
\Text(107,122)[tr]{\bf {\scriptsize 4-2-2}}
\Text(107,53)[br]{\bf {\scriptsize 5-1}}
\Text(110,5)[t]{$e_1$}
\Text(5,125)[r]{$e_2$}
\Text(50,45)[t]{$z=0$}
\Text(50,205)[b]{$z=2\pi i R_6$}
\Text(170,45)[t]{$z=2\pi R_5$}
\Vertex(60,50){3}
\Vertex(160,50){3}
\Vertex(60,200){3}
\Vertex(160,200){3}
\Text(60,50)[c]{$\times$}
\Text(110,50)[c]{$\times$}
\Text(110,125)[c]{$\times$}
\Text(60,125)[c]{$\times$}
\end{picture}
\caption{The $T^2/Z_2$ orbifold in the $z=x_5+ix_6$ plane.  Each
 orbifold fixed point is denoted by a cross and labelled by the
 non-trivial gauge transformations acting on it: 10 for $SO(10)$, 5-1
 for $SU(5) \otimes U(1)_X$, $5'$-$1'$ for $SU(5)' \otimes U(1)'_X$ and
 4-2-2 for $SU(4)_C \otimes SU(2)_L \otimes SU(2)_R$.  The physical
 space may be taken as the two-sided rectangle formed by folding the
 shaded region along the dotted line and then gluing together the
 touching edges.}
\label{fig:torus}
\end{center}
\end{figure}

For simplicity we take a rectangular lattice for the torus, so that
$e_1=2 \pi R_5$ and $e_2=2\pi i R_6$. Consider the rectangle whose
corners' $z$ coordinates are $0$, $\pi i R_6$, $2 \pi R_5$, and 
$2 \pi R_5 + \pi i R_6$.  The physical space may be taken to be the
two-sided rectangle obtained by folding this rectangle in half along the
$x_5=\pi R_5$ line and then gluing together the edges that are touching
one another (see Fig.~\ref{fig:torus}). The orbifold fixed points are
those that, under the $\pi$ rotation, are mapped into points with which
they were already identified under the translation symmetries of the
torus. There are four orbifold fixed points on the space, whose $z$
coordinates are $0$, $\pi R_5$, $\pi i R_6$, and $\pi (R_5+iR_6)$.
This theory could equally well be constructed with non-orthogonal vectors 
$e_1$ and $e_2$, but the KK mode expansions are simplest for the
orthogonal case.

\subsection{Gauge fields in the bulk}

Consider an $SO(10)$ gauge multiplet propagating in this space.  
The fields may be described by a vector and chiral adjoint multiplet of
4D $N=1$ supersymmetry, $(V,\Phi)$. The bulk action is given by
\cite{Arkani-Hamed:2001tb}
\begin{eqnarray}
  S &=& \int d^6 x \Biggl\{ {1 \over 4 k g^2} 
    {\rm Tr} \left[ \int d^2\theta {\cal W}^{\alpha} {\cal W}_{\alpha}
    + {\rm h.c.} \right] \nonumber\\
  && + \int d^4\theta {1 \over k g^2} 
    {\rm Tr} \Biggl[ (\sqrt{2} \partial^\dagger + \Phi^\dagger) 
    e^{-V} (-\sqrt{2} \partial + \Phi) e^{V} + 
    \partial^\dagger e^{-V} \partial e^V  \Biggr] \Biggr\},
\label{eq:action}
\end{eqnarray}
where $V=V^a T^a$, $\Phi=\Phi^a T^a$, ${\rm Tr}[T^aT^b]=k\delta^{ab}$ 
and $\partial=\partial_5 - i \partial_6$.

Under the torus translations we identify
\begin{eqnarray}
  V(z+e_i) &=& T_i V(z) T_i^{-1}, \\
  \Phi(z+e_i) &=& T_i \Phi(z) T_i^{-1},
\end{eqnarray}
with $(T_1,T_2)=(T_{51},T_{5'1'})$.  Under the orbifold $Z_2$ ($\pi$
rotation), we identify
\begin{eqnarray}
  V(-z) &=& Z V(z) Z^{-1},\\
  \Phi(-z) &=& -Z \Phi(z) Z^{-1},
\end{eqnarray}
with $Z = \sigma_0 \otimes I_5$.

Before proceeding, we have to check that these identifications are
consistent.  First, for both $e_1$ and $e_2$, the same spacetime motion
is induced by initially translating and then rotating as by initially
rotating and then performing an inverse translation.  The two sequences
of operations must yield the same net transformation on the fields.
Since the rotation is gauge trivial,  we require $T_1  =  T_1^{-1}$ and
$T_2 = T_2^{-1}$, which are satisfied by $(T_1,T_2)=(T_{51},T_{5'1'})$. 
The other consistency condition arises because performing an $e_1$
translation followed by an $e_2$ translation induces the same spacetime
motion as does $e_2$ followed by $e_1$. Hence, we require 
$[T_1,T_2] = 0$, which is also clearly satisfied.

\subsection{Gauge symmetries at the orbifold fixed points}

At special points on the orbifold, certain gauge transformation
parameters are forced to vanish.  Therefore, the matter content and
interactions located on the fixed point need only respect the gauge
symmetries whose transformation parameters are non-vanishing there
\cite{Hall:2001pg, Hall:2001tn}.  Consider, for instance, the $z=\pi
R_5$ fixed point.  We have $V(z=\pi R_5)=V(z=-\pi R_5)$ from the $Z_2$
rotation and $V(z=\pi R_5)=T_1 V(z=-\pi R_5) T_1^{-1}$ from the $e_1$
translation.  These equations are consistent only if the wavefunction of
every non-$SU(5) \otimes U(1)_X$ gauge field, and every non-$SU(5)
\otimes U(1)_X$ gauge transformation parameter, vanishes at the $z= \pi
R_5$ fixed point.  Thus, interactions at this point need only preserve
$SU(5) \otimes U(1)_X$.  Similarly, interactions at $z=\pi i R_6$ need
only preserve $SU(5)' \otimes U(1)'_X$.  For the $z= \pi (R_5+iR_6)$
fixed point, one must apply both $e_1$ and $e_2$ translations to compare
with the result from performing the $Z_2$ rotation, and one finds 
$V(z=\pi(R_5+iR_6))=T_1 T_2 V(z=-\pi(R_1+iR_2)) T_2^{-1} T_1^{-1}$.
Since $T_1 T_2 =  \sigma_0 \otimes {\rm diag}(1,1,1,-1,-1)$, which
commutes only with the generators listed in Eq.~(\ref{eq:422gen}), one
learns that interactions at this fixed point need only preserve $SU(4)_C
\otimes SU(2)_L\otimes SU(2)_R$.  Finally, at the fixed point at the
origin, there is no restriction on the gauge transformation parameters,
so that this point preserves the full $SO(10)$. The non-trivial gauge
symmetries acting at each fixed point are shown in 
Table~\ref{tab:fixed-Z2}.  At each fixed point, the 4D $N=1$
supersymmetry is preserved.
\begin{table}
\begin{center}
\begin{tabular}{|c|c|} \hline
  $z$     & gauge symmetry 
\\ \hline
  $0$             & $SO(10)$ \\
  $\pi R_5$       & $SU(5) \otimes U(1)_X$ \\
  $\pi i R_6$     & $SU(5)' \otimes U(1)'_X$ \\
  $\pi(R_5+iR_6)$ & $SU(4)_C \otimes SU(2)_L\otimes SU(2)_R$
\\ \hline
\end{tabular}
\end{center}
\caption{Supersymmetry and gauge symmetry on each of the four fixed points.}
\label{tab:fixed-Z2} 
\end{table}

The appearance of the residual $SU(4)_C \otimes SU(2)_L \otimes SU(2)_R$
gauge symmetry at the $z=\pi (R_5+iR_6)$ fixed point suggests that there
is a modified version of this model that is equally viable.  Instead of
taking $T_2=T_{5'1'}$, so that $z=\pi i R_6$ is an $SU(5)' \otimes
U(1)'_X$ preserving fixed point, one could instead choose $T_2=T_{422}$.
Now the $z=\pi i R_6$ fixed point preserves $SU(4)_C \otimes SU(2)_L
\otimes SU(2)_R$, while the $z=\pi (R_5+iR_6)$ fixed point preserves
$SU(5)' \otimes U(1)'_X$.  The unbroken gauge group is $SU(3)_C \otimes
SU(2)_L \otimes U(1)_Y \otimes U(1)_X$ as before. Note that $T_{422}$ is
a $T_{3R}$ rotation by angle $\pi$, so that $SU(5)$ and flipped $SU(5)$
are related by a $T_{3R}$ flip.

\subsection{Doublet-triplet splitting}
\label{subsec:higgs}

One attractive feature of GUT breaking on orbifolds is that natural
doublet-triplet splitting may occur if the Higgs multiplets propagate in
the bulk \cite{Candelas:1985en}, and it does occur naturally in the
present model.  Consider a hypermultiplet 
${\bf H}_{\bf 10}=(H_{\bf 10},H^c_{\bf 10})$. The form of the Lagrangian
forces us to assign opposite parities to $H_{\bf 10}$ and $H^c_{\bf 10}$
under the $Z_2$ rotation.  Without loss of generality, we take 
$H_{\bf 10}(+)$ and $H_{\bf 10}^c(-)$, so that $H_{\bf 10}^c(z=0)$ is
forced to vanish, and only $H_{\bf 10}$ can contain massless zero modes.
How does $H_{\bf 10}$ transform under the $e_1$ and $e_2$ translations?
For the action to be invariant, we need
\begin{equation}
  H_{\bf 10}(z+e_i)=P_i T_i H_{\bf 10}(z).
\label{eq:higgs}
\end{equation}
Here $P_i$ are $\pm 1$: invariance of the action allows these to be
arbitrary phases, but consistency of the transformation properties of
$H_{\bf 10}$ requires $(P_i T_i)^{-1} = P_i T_i$. 

Under $SU(5)$, $H_{\bf 10}$ decomposes as $H_{\bf 5} + 
H_{\barr{\bf 5}}$.  In terms of our previous notation, the $SU(5)$
generators for these representations come from $S_5+A_5$ and $S_5-A_5$,
respectively.  Referring to Eq.~(\ref{eq:51gen}), we conclude that the
$H_{\bf 5}$ and $H_{\barr{\bf 5}}$ contained in $H_{\bf 10}$ are
eigenvectors of $\sigma_2 \otimes I_5$ with eigenvalues $+1$ and $-1$,
respectively.  Hence the components of $H_{\bf 10}$ have parities under
the $e_1$ and $e_2$ translations given by
\begin{equation}
  H_{\bf 10} = 
  \left( \begin{array}{l}
    h_3(P_1,P_2) \\
    h_2(P_1,-P_2) \\
    \barr{h}_3(-P_1,-P_2) \\
    \barr{h}_2(-P_1,P_2) 
  \end{array} \right),
\label{eq:Ti10}
\end{equation}
where $H_{\bf 5} = (h_3,h_2)$ and $H_{\barr{\bf 5}} =
(\barr{h}_3,\barr{h}_2)$. (For flipped $SU(5)$, the alternative
association of doublets with triplets is made $H_{{\bf 5}'} = 
(h_3,\barr{h}_2)$ and $H_{\barr{\bf 5}'} = (\barr{h}_3,h_2)$.)
No matter what choices are made for $P_i$, only one of $h_3, \barr{h}_3,
h_2$ and $\barr{h}_2$ has a zero mode. As in the $SU(5)$ case 
\cite{Kawamura:2001ev}, doublet-triplet splitting is a necessary
consequence of the orbifold gauge symmetry breaking.

Notice that a single ${\bf H}_{\bf 10}$ hypermultiplet in the bulk leads
to a low energy theory that is anomalous under the unbroken gauge
group. It is necessary to introduce combinations of bulk hypermultiplets
such that the collection of zero modes is anomaly free (bulk anomalies
are discussed in the following subsection). Here we restrict
hypermultiplets to be of low dimension, either ${\bf 10}$ or ${\bf
16}$. With this restriction it is interesting that there are only two
combinations involving a ${\bf 10}$ which give vanishing 4D anomaly:  
${\bf H}_{\bf 10}(P_1, P_2)$ and  ${\bf H}'_{\bf 10}(-P_1, -P_2)$, with
$P_i$ of the same (opposite) signs, leading to zero mode triplets
(doublets)
\begin{eqnarray}
  H_{\bf 10}(+,+) + H'_{\bf 10}(-,-) &\rightarrow& h_3 + \barr{h}_3, 
\nonumber \\
  H_{\bf 10}(+,-) + H'_{\bf 10}(-,+) &\rightarrow& h_2 + \barr{h}_2.
\label{eq:AF10}
\end{eqnarray}
Since ${\bf H}_{\bf 10}$ and ${\bf H}'_{\bf 10}$ must have the same 6D
chirality, these fields cannot have a bulk mass term.  However, they can 
have mass term localized on 4D fixed points.  Here we assume a vanishing
brane mass term between ${\bf H}_{\bf 10}$ and ${\bf H}'_{\bf 10}$.

Starting from a ten dimensional hypermultiplet placed in the bulk, the
orbifold breaking of $SO(10) \rightarrow SU(3)_C \otimes SU(2)_L \otimes
U(1)_Y \otimes U(1)_X$ requires an additional ten dimensional
hypermultiplet to cancel 4D anomalies.  Therefore, doublet-triplet
splitting is a necessary consequence of the orbifold breaking, assuming
that brane localized mass terms are absent. To negate it would require
adding both combinations of Eq.~(\ref{eq:AF10}).  Thus, we can identify
the two Higgs doublets as the smallest set of hypermultiplets whose zero
modes yield vanishing 4D anomaly.

\subsection{Bulk anomalies}
\label{subsection:anomalies}

For the present theory to be consistent, we need cancellation of the
anomalies both on the 4D fixed point, which are calculated by
considering the 4D anomaly of the zero modes, and in the 6D
bulk \cite{Hebecker:2001jb}. Interestingly, the 6D irreducible gauge
anomalies do cancel in the present model with a vector multiplet and two
${\bf 10}$ hypermultiplets ${\bf H}_{\bf 10}$ and ${\bf H}_{\bf 10}'$ in
the bulk \cite{Hebecker:2001jb}.  Moreover, the irreducible gauge
anomaly from a ${\bf 10}$ is equal and opposite to that of a ${\bf 16}$
(or a $\barr{\bf 16}$) with the same 6D chirality, so this cancellation
is maintained provided that each additional ${\bf 10}$ comes with either
a ${\bf 16}$ or a $\barr{\bf 16}$.

There is another irreducible piece in the pure gravitational anomaly, 
but we can cancel it by adding $SO(10)$ singlet fields and/or 
additional hidden gauge groups.  The rest of the anomalies are 
reducible and will be canceled by the Green-Schwarz mechanism 
\cite{Green:1984sg}, leading to axion-like degrees of freedom 
in the low-energy 4D theory.

\subsection{Quarks and leptons in the bulk}

Where should the standard model quarks and leptons be located in this
model?  We first consider the possibility that they appear in
hypermultiplet ${\bf 16}$s that propagate in the bulk.  As discussed
above, we are then led to introduce an extra  ${\bf 10}$ for each
${\bf 16}$ to cancel irreducible gauge anomalies.  We assume that these
extra ${\bf 10}$s pair up to become heavy through large brane localized
mass terms (although the ${\bf 10}$s containing the Higgs doublets must
not obtain such a mass term).

\subsubsection{Minimal matter content from anomaly cancellation}

Imagine a hypermultiplet $\Psi_{\bf 16}=(\psi_{\bf 16},
\psi^c_{\barr{\bf 16}})$ propagating in the bulk.  We assign the
parities under the $Z_2$ rotation as $(+,-)$ so that the conjugate
matter does not have a zero mode.  The transformation properties of
$\psi_{\bf 16}$ under the torus translations are most easily deduced by
considering the $SU(5) \otimes U(1)_X$ decomposition of the matter
couplings to gauge fields.  The gauge fields decompose as ${\bf 45} =
{\bf 24}_0 + {\bf 10}_4 + \barr{\bf 10}_{-4} + {\bf 1}_0$, and the
matter decomposes as ${\bf 16} = {\bf 10}_{-1} + \barr{\bf 5}_3 + 
{\bf 1}_{-5}$.  Since we know that the ${\bf 10}_4$ and 
$\barr{\bf 10}_{-4}$ gauge fields have negative parity under the $e_1$
translation, the existence of the $\psi_{\barr{\bf 5}}^{\dagger} 
V_{\bf 10} \psi_{\bf 10}$ and $\psi_{\bf 10}^{\dagger} V_{\bf 10} 
\psi_{\bf 1}$ terms in the Lagrangian implies that we can take the
parities for $(\psi_{\bf 10}, \psi_{\barr{\bf 5}}, \psi_{\bf 1})$ to be
either $(-,+,+)$ or $(+,-,-)$ ($U(1)_X$ charges omitted for notational
simplicity).  Similarly, under the $e_2$ translation one can take the
parities for $(\psi_{{\bf 10}'}, \psi_{\barr{\bf 5}'}, \psi_{{\bf 1}'})$
to be either $(-,+,+)$ or $(+,-,-)$.  Here ${\bf 10}'_{-1}$, 
$\barr{\bf 5}'_3$ and ${\bf 1}'_{-5}$ denote the decomposition of 
${\bf 16}$ under $SU(5)'$, or flipped $SU(5)$, which means that 
$\psi_{{\bf 10}'}=(Q, D, N)$, $\psi_{\barr{\bf 5}'}=(U, L)$ and
$\psi_{{\bf 1}'}=E$.  Overall, there are four choices for the parities:
\begin{eqnarray}
  {\bf 16}_{++}&~~~&\mbox{(zero mode)~}=Q, \nonumber\\
  {\bf 16}_{+-}&~~~&\mbox{(zero modes)}=U,E, \nonumber\\
  {\bf 16}_{-+}&~~~&\mbox{(zero modes)}=D,N, \\
  {\bf 16}_{--}&~~~&\mbox{(zero mode)~}=L, \nonumber  
\end{eqnarray}
where the first $\pm$ sign refers to the parity of ${\bf 10}_{-1}$ under 
the $e_1$ translation, while the second refers to the parity of 
${\bf 10}'_{-1}$ under the $e_2$.  Thus, for each generation of matter
that propagates in the bulk, we need four hypermultiplets whose parities
under the torus translations conspire to yield a complete $SO(10)$
multiplet for the massless zero modes.

The conspiracy of parities required to obtain a complete generation of
massless zero modes is required by cancellation of zero mode anomalies.
In subsection \ref{subsec:higgs}, we saw that the smallest anomaly-free
set came from two ${\bf 10}$s, forming two Higgs doublets.  We now show
that the smallest anomaly-free set of hypermultiplets with {\it chiral}
zero modes under $SU(3)_C \otimes SU(2)_L \otimes U(1)_Y \otimes U(1)_X$ 
is the combination of four ${\bf 16}$s described above, forming a single
generation of matter including the right-handed neutrino.

The smallest representation of $SO(10)$ is the vector, ${\bf 10}$, but
since this representation is real, we cannot obtain chiral matter from
any combination of ${\bf 10}$s.  The next simplest possibility is to try
some combination of ${\bf 10}$s and ${\bf 16}$s.  If there were no
$U(1)_X$, the smallest anomaly free combination which is chiral would be 
${\bf 10}_{-+},~{\bf 10}_{--},~{\bf 16}_{++}$ and ${\bf 16}_{+-}$,
leaving $L,~D,~Q$ and $\{U, E\}$, respectively, in zero modes.  However, 
this set is anomalous under $U(1)_X$.  Remember that coming from 
${\bf 10}$s, $L$ and $D$ have a `wrong' $U(1)_X$ quantum number of $-2$.
The simplest potential cure would be to add the right-handed neutrino
$N$.  Note that ${\bf 10}$ cannot give $N$, so that we must use 
${\bf 16}_{-+}$, which gives $D$ and $N$, instead of ${\bf 10}_{--}$.  
Even then, however, the wrong $U(1)_X$ charge of $L$ from 
${\bf 10}_{-+}$ still fails to cancel anomaly associated with $U(1)_X$,
forcing us to give up ${\bf 10}_{-+}$ and to use $\bf 16_{--}$.

Therefore, although four ${\bf 16}$ hypermultiplets are needed to
complete a single generation, it does not spoil much the unification of
matter as a virtue of underlying $SO(10)$.  One might think that if we
identify matter as the smallest set of hypermultiplets giving
anomaly-free chiral zero modes, then $SU(5)$ can also explain the matter 
quantum numbers.  In the $SO(10)$ case, however, there is naturally a
$U(1)_X$ gauge symmetry, so that it requires the right-handed neutrino 
$N$ to be present in a generation.

The matter fields can couple to the Higgs fields on any one of the four
fixed points.  For instance, on the $SO(10)$ preserving brane we have
the Yukawa couplings
\begin{equation}
  S = \int d^6 x \delta^2(z) \left\{ 
    \int d^2\theta \left(\lambda \psi_{\bf 16} \psi_{\bf 16} H_{\bf 10} 
    + {\lambda}' \psi_{\bf 16} \psi_{\bf 16} H_{\bf 10}' \right)
    + {\rm h.c.} \right\},
\end{equation}
where we have suppressed indices labeling the various $\psi_{\bf 16}$'s.
Note that there are no GUT fermion mass relations in the present setup
because the massless zero modes that comprise a single generation
originate from different $\bf 16$s.

\subsubsection{Supersymmetry breaking, $U(1)_X$ breaking, and neutrino
masses}

What is an appropriate mechanism for breaking the remaining 4D $N=1$
supersymmetry?  Gaugino mediation is not accommodated by the present
model because matter propagates in the bulk and cannot be spatially
separated from the supersymmetry breaking.  A different, feasible
mechanism, which we can adopt here, is given by the type of
Scherk-Schwarz supersymmetry breaking described in
Ref.~\cite{Barbieri:2001yz}, involving a small parameter that lowers the 
breaking scale relative to the compactification scale.

Two other phenomenological questions concern the unbroken $U(1)_X$ and
the origin of masses for the right-handed neutrinos.  We could try to
break $U(1)_X$ by more complicated orbifolds, where parity operations
become noncommutative, but we will not investigate this possibility
here.  Instead, we break $U(1)_X$ by driving a GUT-scale vacuum
expectation value (VEV) for a field $X$ transforming as a singlet under
$SU(5)$ but with charge 10 under $U(1)_X$.  (The $\barr{X}$ field with
the opposite $U(1)_X$ charge is also introduced to cancel anomalies.)
This VEV can also be used to give a large mass to the right-handed
neutrinos.  Since the field is an $SU(5)$ singlet, we must localize it
to the $SU(5) \otimes U(1)_X$ brane.  The interaction term responsible
for $U(1)_X$ breaking is
\begin{equation}
  S = \int d^6 x \delta^2(z-\pi R_5) \left\{ 
    \int d^2\theta k X N_{\psi} N_{\psi} + {\rm h.c.} \right\},
\label{eq:snn}
\end{equation}
where $N_\psi$ is the right-handed neutrino coming from the bulk
hypermultiplet $\psi_{\bf 16}$.

If Scherk-Schwarz supersymmetry breaking is employed as in 
Ref.~\cite{Barbieri:2001yz}, it gives soft supersymmetry breaking masses
of the order the weak scale to $N_{\psi}$, while $X$ remains
massless, since it is located on the brane.  Therefore, the above
interaction drives the mass-squared of $X$ scalar negative, while that
of $N_{\psi}$ remains positive.  Since there is no large quartic
potential in the flat direction $X = \barr{X}$, we have a runaway
situation and obtain a huge VEV for $X$ (presumably around the
compactification scale), which breaks $U(1)_X$ at very high energy
scale.

After this breaking, the right-handed neutrinos receive large Majorana
masses from Eq.~(\ref{eq:snn}).  In order to get the right order of
magnitude for the neutrino masses through the see-saw mechanism, the $k$ 
couplings must be somewhat small of order $10^{-2}-10^{-3}$.

\subsection{Quarks and leptons on branes}
\label{subsec:qlb}

Now we consider an alternative possibility that the quarks and leptons
are localized to one of the fixed points.  If they live on the $z=0$
fixed point the quarks and leptons are forced to appear in full $SO(10)$ 
multiplets, since the full $SO(10)$ gauge symmetry is realized there.
We thus introduce three $\psi_{\bf 16}$'s each transforming as 
${\bf 16}$ under $SO(10)$.  This setup provides the same understanding
of the standard-model fermion quantum numbers as is given by the
standard 4D $SO(10)$.

Brane-localized interactions of the ${\psi_{\bf 16}}$'s with the bulk
Higgs multiplets give rise to Yukawa couplings
\begin{equation}
  S = \int d^6 x \delta^2(z) \left\{ \int d^2\theta \left( 
    \lambda \psi_{\bf 16} \psi_{\bf 16} H_{\bf 10} 
    + \lambda' \psi_{\bf 16} \psi_{\bf 16} H'_{\bf 10} \right) 
    + {\rm h.c.} \right\}.
\label{yukawa}
\end{equation}
Since the up- and down-type Higgs doublets come from different $SO(10)$
multiplets, the fermion mass relations are those of $SU(5)$ rather than
those of $SO(10)$.  Realistic fermion masses may be obtained through
mixing of the $\psi_{\bf 16}$'s on the fixed point with bulk 
${\bf 16}$s, in a way similar to what was done for fermion masses in the
$SU(5)$ case in Ref.~\cite{Hall:2001pg}.  Again, for cancellation of
irreducible gauge anomalies, these bulk ${\bf 16}$s must be accompanied
by ${\bf 10}$ hypermultiplets, which must obtain large brane localized
mass terms.

If matter is localized to one of the other fixed points, its Yukawa
interactions will only respect the reduced gauge symmetry remaining at
the fixed point.  On the 5-1 and 4-2-2 branes, these interactions give
rise to $SU(5)$ fermion mass relations, which again may be corrected
through mixing with bulk states.  On the $5'$-$1'$ fixed point, on the
other hand, there are no GUT relations for the fermion masses, other
than equality between the up-type quark and neutrino Dirac mass
matrices.

With matter localized on one of the fixed points, gaugino mediation of
supersymmetry breaking \cite{Kaplan:2000ac} is easily accommodated,
provided that supersymmetry breaking occurs on a different fixed
point from the one where matter resides.  Gaugino mass relations at
the compactification scale depend on which is the supersymmetry breaking
fixed point: we have $M_3=M_2=M_1$ if supersymmetry is broken on the
10 or 5-1 branes, and $M_3 = M_2 \neq M_1$ if supersymmetry is broken on
the 4-2-2 or $5'$-$1'$ branes.

As in the bulk matter case, an obvious location for $U(1)_X$ breaking is
the 5-1 brane.  However, unless matter is localized to this brane, it
does not feel the breaking directly, so that communication through
exchanges of bulk states is required.  Consider, for example, the case
where matter lives on the 10 brane.  Suppose that bulk states 
$\chi_{\bf 16} + \barr{\chi}_{\barr{\bf 16}}$ with a brane mass term
couple to the 5-1 and 10 branes according to
\begin{eqnarray}
  S &=& \int d^2\theta \Biggl\{ \delta^2(z-\pi R_5) 
    \left( Y (X \barr{X} - \mu^2 )
    + X N_{\barr{\chi}} N_{\barr{\chi}}
    + \barr{X} \barr{N}_{\chi} \barr{N}_{\chi} \right)
\nonumber\\
  && \qquad + \delta^2(z) \barr{\chi}_{\barr{\bf 16}} 
    \barr{\chi}_{\barr{\bf 16}} \psi_{\bf 16} \psi_{\bf 16} \Biggr\}
    + {\rm h.c.}
\end{eqnarray}
Here we have neglected coupling constants, $Y$ is a singlet superfield,
and $X$ and $\barr{X}$ are $SU(5)$ singlets with $U(1)_X$ charges $10$
and $-10$, respectively.  $N_{\barr{\chi}}$ represents ``right-handed 
neutrino'' components in $\barr{\chi}_{\barr{\bf 16}}$ and similarly for 
$\barr{N}_{\chi}$.  The first term in the above interaction forces $X$
and $\barr{X}$ to acquire VEVs equal to $\mu$. Upon integrating out 
$N_{\barr{\chi}}$ and $\barr{N}_{\chi}$ states, the non-local term $X
N_{\psi} N_{\psi}$ is generated, giving rise to masses for the
right-handed neutrinos $N_{\psi}$.  This term carries a suppression by
powers of $(M_\chi R)$ if the brane mass term $M_\chi$ for 
$\chi_{\bf 16}+\barr{\chi}_{\barr{\bf 16}}$ is localized on $z = 0$ or
$\pi R_5$, since the wavefunctions for $\chi_{\bf 16}$ and 
$\barr{\chi}_{\barr{\bf 16}}$ are suppressed there.  This may give the
correct order of magnitude for the right-handed neutrino masses.

\section{A Model on $T^2/Z_6$}
\label{section:t2z6}

\subsection{Orbifold structure}

As discussed in subsection \ref{subsection:anomalies}, the $T^2/Z_2$
model requires the Green-Schwarz mechanism for anomaly cancellation,
which leads to axion-like degrees of freedom in the low energy theory.
Here we consider a different setup for
$SO(10)$ in 6D, with $N=2$ supersymmetry,
so that all bulk anomalies automatically cancel.
The $N=2$ supersymmetry in 6D corresponds to $N=4$ supersymmetry in 4D,
so that only the gauge multiplet can be introduced in the bulk.  This
multiplet can be decomposed under a 4D $N=1$ supersymmetry into a vector
multiplet $V$ and three chiral multiplets $\Sigma$, $\Phi$, and $\Phi^c$
in the adjoint representation, with the fifth and sixth components of the gauge
field, $A_5$ and $A_6$, contained in the lowest component of $\Sigma$.

Using 4D $N=1$ language, the bulk action may be written as
\cite{Arkani-Hamed:2001tb}
\begin{eqnarray}
  S &=& \int d^6 x \Biggl\{
  {\rm Tr} \Biggl[ \int d^2\theta \left( \frac{1}{4 k g^2} 
  {\cal W}^\alpha {\cal W}_\alpha + \frac{1}{k g^2} 
  \left( \Phi^c \partial \Phi   - \frac{1}{\sqrt{2}} \Sigma 
  [\Phi, \Phi^c] \right) \right) + {\rm h.c.} \Biggr] 
\nonumber\\
  && + \int d^4\theta \frac{1}{k g^2} {\rm Tr} \Biggl[ 
  (\sqrt{2} \partial^\dagger + \Sigma^\dagger) e^{-V} 
  (-\sqrt{2} \partial + \Sigma) e^{V}
  + \Phi^\dagger e^{-V} \Phi  e^{V}
  + {\Phi^c}^\dagger e^{-V} \Phi^c e^{V} 
\Biggr] \Biggr\},
\label{eq:t2z6action}
\end{eqnarray}
in the Wess-Zumino gauge.

Can we build a realistic model starting with 6D $N=2$ supersymmetry on
$T^2/Z_2$?  The trouble is that for this orbifold the fixed points are
left with 4D $N=2$ supersymmetry rather than 4D $N=1$ supersymmetry.  To reduce
the supersymmetry further requires an orbifold in which more modding out
is done --- a fairly simple orbifold that works is $T^2/Z_6$.  This
orbifold is constructed by identifying points of the infinite plane
$R^2$ under three operations, ${\cal Z}: z \rightarrow \omega z$, 
${\cal T}_1: z \rightarrow z+2\pi R$ and 
${\cal T}_2: z \rightarrow z+2\pi \omega R$, where 
$\omega = e^{i\pi/3}$.  The identifications for the fields under 
${\cal Z}$ are taken to be 
\begin{eqnarray}
  V(\omega z) &=& T_{422} V(z) T_{422}^{-1},
\\
  \Sigma(\omega z) &=& \omega^5 T_{422} \Sigma(z) T_{422}^{-1},
\\
  \Phi(\omega z) &=& \omega^5 T_{422} \Phi(z) T_{422}^{-1},
\\
  \Phi^c(\omega z) &=& \omega^2 T_{422} \Phi^c(z) T_{422}^{-1},
\end{eqnarray}
and the identifications under ${\cal T}_1$ and ${\cal T}_2$ are
\begin{eqnarray}
  V(z+2\pi R) &=& T_{51} V(z) T_{51}^{-1},
\\
  \Sigma (z+2\pi R) &=& T_{51} \Sigma (z) T_{51}^{-1},
\\
  \Phi(z+2\pi R) &=& T_{51} \Phi (z) T_{51}^{-1},
\\
  \Phi^c(z+2\pi R) &=& T_{51} \Phi^c(z) T_{51}^{-1},
\end{eqnarray}
and 
\begin{eqnarray}
  V(z+2\pi  \omega R) &=&T_{51} V(z) T_{51}^{-1},
\\
  \Sigma (z+2\pi  \omega R) &=& T_{51} \Sigma (z) T_{51}^{-1},
\\
  \Phi(z+2\pi  \omega  R) &=& T_{51} \Phi (z) T_{51}^{-1},
\\
  \Phi^c(z+2\pi  \omega  R) &=& T_{51} \Phi^c(z) T_{51}^{-1},
\end{eqnarray}
respectively.  This choice of identifications breaks the $SO(10)$ gauge
group to $SU(3)_C \otimes SU(2)_L \otimes U(1)_Y \otimes U(1)_X$ at low
energies, and the only massless zero modes are those of $V$.  This
orbifold has a single fixed point located at $z=0$, which has 4D $N=1$
supersymmetry and 4-2-2 gauge symmetry.  However, there is another
special point that is fixed under the $Z_3$ subgroup of $Z_6$, located
at $z=(2 \pi R/\sqrt{3}) e^{i \pi /6}$.  This point has 5-1 gauge
symmetry, and also has only 4D $N=1$ supersymmetry. 

\subsection{Matter configuration}

There are only two possibilities for where the quarks and leptons live
in this model, corresponding to the 4-2-2 and 5-1 points. Higgs
multiplets are not allowed to propagate in the bulk because of the 6D
$N=2$ supersymmetry, so we are forced to put the Higgs on the same brane
as the quarks and leptons.  If we choose the 5-1 brane, we are faced
with a difficult doublet-triplet splitting problem, as the Higgs
doublets must appear in ${\bf 5}_{2} + \barr{\bf 5}_{-2}$ multiplets
under $SU(5) \otimes U(1)_X$, and the colored components must somehow
get heavy.  We thus focus on the alternative placement on the 4-2-2
brane.  On this point a generation of matter is formed from the $SU(4)_C
\otimes SU(2)_L \otimes SU(2)_R$ multiplets $({\bf 4},{\bf 2},{\bf 1}) + 
(\barr{\bf 4},{\bf 1}, {\bf 2})$, and the Higgs doublets of the MSSM are
contained in a $({\bf 1},{\bf 2},{\bf 2})$ multiplet.  Gaugino mediation
is naturally realized in this model by localizing the breaking of 4D
$N=1$ supersymmetry to the 5-1 brane.

\subsection{$U(1)_X$ breaking and fermion masses}

The $U(1)_X$ is easily broken on the 5-1 brane by introducing a
brane-localized $SU(5)$ singlet $X$ charged under $U(1)_X$, and a
superpotential that forces it to acquire a VEV.  However, the question
of how this $U(1)_X$ breaking is communicated to the 4-2-2 brane to give
rise to Majorana masses for the right-handed neutrinos is not
straightforward.  The states contained in the adjoint of
$SO(10)$ do not have the correct quantum numbers to generate $XNN$ as a
non-local operator, and the 6D $N=2$ supersymmetry prevents us from
adding additional bulk states.  A problem related to that of the
right-handed neutrino masses is that this model has $SO(10)$ fermion
mass relations: somehow, the 4-2-2 brane must be made to feel 4-2-2
breaking.

Right-handed neutrino masses could be generated if a multiplet $X'$ on
the 4-2-2 brane acquired a VEV that broke $U(1)_X$ but not the standard
model gauge group.  In this case, however, there is a vacuum alignment
problem because the potential for $X'$ is 4-2-2 symmetric.
Correspondingly, a potential that forces $X'$ to take on a VEV will lead 
to extra massless Goldstone states.  For instance, such a VEV cannot be
$SU(2)_R$ globally symmetric, and there are no $SU(2)_R$ gauge bosons to
eat the Goldstones, as they are already made heavy by the orbifold
compactification.

Changes to this picture come from radiative corrections to the potential
for $X'$ below the compactification scale.  If the theory were not
supersymmetric these corrections would give masses of order 
$\alpha v/(4 \pi)$ to the Goldstones, where $v$ is the VEV of $X'$.
However, in the supersymmetric limit the Goldstones pick up no mass, and
they thus only acquire TeV-scale masses from gaugino mediation, just as
do the squarks and sleptons.  Whether it is a realistic possibility that
these corrections to the potential force $X'$ to point in an appropriate
direction is a question we leave for future study.  In any case this
setup reveals a crucial point: in theories where a gauge generator is
broken both by the orbifold projection and by a brane VEV, there will be
a corresponding ``would-be Goldstone'' with mass $\sim m_{\rm SUSY}$.
These states generically spoil the success of the gauge coupling
unification in the MSSM, and might be problematic for proton
stability.   This is independent of $1/R$ and the scale of
the brane VEV. There is such a TeV supermultiplet for each generator
which is ``broken twice''.

\section{A Model on $T^2/(Z_2 \times Z_2')$}
\label{section:z2sq} 

In the $T^2/Z_2$ model of section \ref{section:z2}, the $U(1)_X$ left
over after orbifolding was broken by the VEV of a field transforming
under $U(1)_X$ only. In contrast, in the model of section
\ref{section:t2z6} either we are left with no Majorana masses for the
right-handed neutrinos, or we must require a field $X'$ to have
additional gauge transformation properties and a vacuum alignment
problem must be resolved to ensure that the standard model gauge group
remains unbroken at low energies.  In this section we construct a third
model, on a $T^2/(Z_2 \times Z_2')$ orbifold, in which both the vacuum
alignment problem of the $T^2/Z_6$ model, as well as the anomalies of
the $T^2/Z_2$ model, are absent.

\subsection{Orbifold structure}

We again consider a theory with 6D $N=2$ supersymmetry. Using the 4D
$N=1$ language, we can express the bulk action as
\cite{Arkani-Hamed:2001tb}
\begin{eqnarray}
  S &=& \int d^6 x \Biggl\{
  {\rm Tr} \Biggl[ \int d^2\theta \left( \frac{1}{4 k g^2} 
  {\cal W}^\alpha {\cal W}_\alpha + \frac{1}{k g^2} 
  \left( \Phi \partial_5 \Sigma_6 - \Phi \partial_6 \Sigma_5
  - \frac{1}{\sqrt{2}} \Phi 
  [\Sigma_5, \Sigma_6] \right) \right) + {\rm h.c.} \Biggr] 
\nonumber\\
  && + \int d^4\theta \frac{1}{k g^2} {\rm Tr} \Biggl[ 
  (\sqrt{2} \partial_5 + \Sigma_5^\dagger) e^{-V} 
  (-\sqrt{2} \partial_5 + \Sigma_5) e^{V}
  + (\sqrt{2} \partial_6 + \Sigma_6^\dagger) e^{-V} 
  (-\sqrt{2} \partial_6 + \Sigma_6) e^{V}
\nonumber\\
  && \qquad \qquad \qquad
  + \Phi^\dagger e^{-V} \Phi e^{V} 
  + \partial_5 e^{-V} \partial_5 e^{V}
  + \partial_6 e^{-V} \partial_6 e^{V} \Biggr] \Biggr\},
\label{eq:5daction}
\end{eqnarray}
in the Wess-Zumino gauge.   When expressed in terms of components, this
action and that of Eq.~(\ref{eq:t2z6action}) have identical forms.  The
orbifold of the present model will preserve a different 4D $N=1$
supersymmetry than the orbifold of the previous one (namely, one in
which $A_5$ and $A_6$ appear in different superfields), and we have
chosen to make this different 4D $N=1$ supersymmetry manifest. 

The orbifold $T^2/(Z_2 \times Z_2')$ is constructed by identifying
points of the infinite plane $R^2$ under four operations, 
${\cal Z}_1: (x^5,x^6) \rightarrow (-x^5,x^6)$, 
${\cal Z}_2: (x^5,x^6) \rightarrow (x^5,-x^6)$, 
${\cal T}_1: (x^5,x^6) \rightarrow (x^5+2\pi R_5,x^6)$ and 
${\cal T}_2: (x^5,x^6) \rightarrow (x^5,x^6+2\pi R_6)$.  Here, for
simplicity, we have taken the two translations ${\cal T}_1$ and 
${\cal T}_2$ to be in orthogonal directions.

Under ${\cal Z}_1$ and ${\cal Z}_2$ we make the gauge-trivial
identifications
\begin{eqnarray}
  V(-x^5,x^6) &=& V(x^5,x^6) ,
\label{eq:z1-1} \\
  \Sigma_5(-x^5,x^6) &=& -  \Sigma_5(x^5,x^6),
\label{eq:z1-2} \\
  \Sigma_6(-x^5,x^6) &=& \Sigma_6(x^5,x^6),
\label{eq:z1-3} \\
  \Phi(-x^5,x^6) &=& -\Phi(x^5,x^6),
\label{eq:z1-4} 
\end{eqnarray}
and
\begin{eqnarray}
  V(x^5,-x^6) &=& V(x^5,x^6),
\label{eq:z2-1} \\
  \Sigma_5(x^5,-x^6) &=&  \Sigma_5(x^5,x^6),
\label{eq:z2-2} \\
  \Sigma_6(x^5,-x^6) &=& - \Sigma_6(x^5,x^6),
\label{eq:z2-3} \\
  \Phi(x^5,-x^6) &=& - \Phi(x^5,x^6),
\label{eq:z2-4} 
\end{eqnarray}
respectively. Note that various signs appearing in 
Eqs.~(\ref{eq:z1-1} -- \ref{eq:z2-4}) are determined by invariance of
the bulk action under the ${\cal Z}_{1,2}$ operations.

The ${\cal Z}_1$ identification breaks 4D $N=4$ supersymmetry to 4D
$N=2$ supersymmetry (or equivalently, 6D $N=2$ to 6D $N=1$
supersymmetry), with $(V,\Sigma_6)$ forming a vector multiplet and
$(\Sigma_5,\Phi)$ forming a hypermultiplet.  Similarly, the 
${\cal Z}_2$ identification breaks 4D $N=4$ supersymmetry to 4D $N=2$
supersymmetry, with $(V,\Sigma_5)$ forming a vector multiplet and
$(\Sigma_6,\Phi)$ forming a hypermultiplet.  This means that the two
$N=2$ supersymmetries remaining after the ${\cal Z}_1$ and ${\cal Z}_2$ 
operations are different subgroups of the original $N=4$ supersymmetry.
Thus, the combination of ${\cal Z}_1$ and ${\cal Z}_2$ identifications, 
{\it i.e.} the $T^2/(Z_2 \times Z_2')$ compactification, breaks the 
original 6D $N=2$ supersymmetry all the way down to 4D $N=1$ 
supersymmetry.

The ${\cal T}_1$ and ${\cal T}_2$ identifications are
\begin{eqnarray}
  V(x^5+2\pi R_5,x^6) &=& T_{51} V(x^5,x^6) T_{51}^{-1},
\label{eq:t5-1} \\
  \Sigma_5(x^5+2\pi R_5,x^6) &=& T_{51} \Sigma_5(x^5,x^6) T_{51}^{-1},
\label{eq:t5-2} \\
  \Sigma_6(x^5+2\pi R_5,x^6) &=& T_{51} \Sigma_6(x^5,x^6) T_{51}^{-1},
\label{eq:t5-3} \\
  \Phi(x^5+2\pi R_5,x^6) &=& T_{51} \Phi(x^5,x^6) T_{51}^{-1},
\label{eq:t5-4} 
\end{eqnarray}
and 
\begin{eqnarray}
  V(x^5,x^6+2\pi R_6) &=& T_{422} V(x^5,x^6) T_{422}^{-1},
\label{eq:t6-1} \\
  \Sigma_5(x^5,x^6+2\pi R_6) &=& T_{422} \Sigma_5(x^5,x^6) T_{422}^{-1},
\label{eq:t6-2} \\
  \Sigma_6(x^5,x^6+2\pi R_6) &=& T_{422} \Sigma_6(x^5,x^6) T_{422}^{-1},
\label{eq:t6-3} \\
  \Phi(x^5,x^6+2\pi R_6) &=& T_{422} \Phi(x^5,x^6) T_{422}^{-1},
\label{eq:t6-4} 
\end{eqnarray}
respectively.  These identifications leave the 3-2-1-1 components of $V$
as the only ones with massless zero modes.  (We could chose 
$(T_{51},T_{5'1'})$ or $(T_{5'1'},T_{422})$, instead of
$(T_{51},T_{422})$, for $({\cal T}_1,{\cal T}_2)$ operations.  All the
arguments in the rest of this section can be extended to these cases in
a straightforward way.)

The structure of the fixed points can be worked out by considering the 
profiles of symmetry transformation parameters in the extra dimensions.
On each of the four fixed points of the $T^2/(Z_2 \times Z_2')$ orbifold, 
the remaining supersymmetry and gauge symmetry is given in 
Table~\ref{tab:fixedpoints} ---
\begin{table}
\begin{center}
\begin{tabular}{|c|c|c|} \hline
  $(x^5, x^6)$     & 4D supersymmetry & gauge symmetry 
\\ \hline
  $(0,   0)$           & $N=1$ & $SO(10)$ \\
  $(\pi R_5, 0)$       & $N=1$ & $SU(5)\otimes U(1)_X$ \\
  $(0, \pi R_6)$       & $N=1$ & $SU(4)_C \otimes SU(2)_L \otimes SU(2)_R$ \\
  $(\pi R_5, \pi R_6)$ & $N=1$ & $SU(3)_C \otimes 
                                  SU(2)_L \otimes U(1)_Y \otimes U(1)_X$ 
\\ \hline
\end{tabular}
\end{center}
\caption{Supersymmetry and gauge symmetry on each of the four fixed points.}
\label{tab:fixedpoints} 
\end{table}
matter multiplets and interactions placed on the fixed points must
respect these symmetries.  An important feature of this orbifold is that
these fixed points are connected by ``fixed lines'' with reduced
supersymmetry and gauge symmetry.  These lines are fixed with respect to
one of the $Z_2$ reflections but not the other.  The remaining
supersymmetry and gauge symmetry for each such line are given in
Table~\ref{tab:fixedlines}.
\begin{table}
\begin{center}
\begin{tabular}{|c|c|c|} \hline
  fixed lines      & 4D supersymmetry & gauge symmetry 
\\ \hline
  $x^5 = 0$       & $N=2$ & $SO(10)$ \\
  $x^6 = 0$       & $N=2$ & $SO(10)$ \\
  $x^5 = \pi R_5$ & $N=2$ & $SU(5) \otimes U(1)_X$ \\
  $x^6 = \pi R_6$ & $N=2$ & $SU(4)_C \otimes SU(2)_L \otimes SU(2)_R$ 
\\ \hline
\end{tabular}
\end{center}
\caption{Supersymmetry and gauge symmetry on each of the four fixed lines.}
\label{tab:fixedlines} 
\end{table}
Because of the reduced supersymmetry on these lines,
we have additional $(4+1)$-dimensional subspaces on which matter
multiplets may be placed, without giving rise to anomalies of the 6D
bulk. The rich fixed point and fixed line structure of this orbifold
provides for a multitude of possibilities for matter locations,
fermion mass relations and $U(1)_X$ breaking, some of which we briefly
describe in the next subsection.

\subsection{Matter configurations}

The quarks and leptons may reside on any of the four fixed points or on
any of the four fixed lines.  If they are localized to the fixed points,
for example, a single generation arises as a ${\bf 16}$ of $SO(10)$ for
the 10 brane, as ${\bf 10}_{-1} + \barr{\bf 5}_{3} + {\bf 1}_{-5}$ under 
$SU(5) \otimes U(1)_X$ for the 5-1 brane, as $({\bf 4},{\bf 2},{\bf 1})
+ (\barr{\bf 4},{\bf 1},{\bf 2})$ under $SU(4)_C \otimes SU(2)_L \otimes
SU(2)_R$ for the 4-2-2 brane, and as the matter multiplets of the
standard model, with appropriate $U(1)_X$ charges, on the 3-2-1-1 brane.
Wherever matter resides, the Higgs multiplets should live on a fixed
line or point that is in contact with the matter in order to give rise
to Yukawa couplings.

For the Yukawa couplings to be on either the 4-2-2 or 3-2-1-1 branes the
Higgs multiplets can either propagate on a touching fixed line, or they
can live on those points as a $({\bf 1},{\bf 2},{\bf 2})$ or 
$({\bf 1},{\bf 2},1/2,2) + ({\bf 1},{\bf 2},-1/2,-2)$ under the residual
4-2-2 and 3-2-1-1 gauge symmetries, respectively.  If they propagate on
either the $x^5=0$ $SO(10)$ line or the $x^5=\pi R_5$ 5-1 line, natural
doublet-triplet splitting arises by assigning parities so that the
colored triplet zero modes are projected out (this will be illustrated
in a specific example shortly).  If they propagate on the 4-2-2 line,
the colored triplets can be avoided from the start by introducing only
$({\bf 1},{\bf 2},{\bf 2})$ multiplets on the line.

If, on the other hand, the Yukawa couplings are on either the 10 or 5-1
branes it is advantageous for the Higgs multiplets to propagate on the
$x^5=0$ $SO(10)$ line or the $x^5=\pi R_5$ 5-1 line, respectively.  The
reason is that otherwise the Higgs fields are spatially separated from
the $SU(5)$ breaking that arises from orbifolding, making
doublet-triplet splitting more problematic.

Gaugino mediated supersymmetry breaking is again easily accommodated by
this orbifold, by localizing the supersymmetry breaking to a fixed point
from which matter is spatially separated \cite{Hall:2001zb}.  All three
gaugino masses unify if the supersymmetry breaking is on either the 10
or 5-1 points, but there is no unification if the supersymmetry breaking
is on either the 4-2-2 or 3-2-1-1 points. 

We require $U(1)_X$ to be broken by the VEV of $SU(5)$ singlet fields
$X$ and $\barr{X}$, with $U(1)_X$ charges $10$ and $-10$, which
therefore live on either the 5-1 or 3-2-1-1 points.  These fields
acquire equal VEVs through the brane-localized superpotential
$Y(X \barr{X}-\mu^2)$.  If the matter fields propagate on a line
touching the point where $U(1)_X$ is broken, the right-handed neutrinos
obtain masses through the direct superpotential coupling $XNN$.
Otherwise, the breaking must be communicated by heavy states propagating
on the fixed lines.

Clearly, there are numerous interesting theories that may be built on
this orbifold.  Here we simply consider two simple illustrative
examples.  Suppose that quarks and leptons are contained in three
$\psi_{\bf 16}$'s that live on the $SO(10)$ fixed point.  The best
choice for the Higgs multiplets is for them to be contained in a
hypermultiplet ${\bf H}_{\bf 10}$ that propagates on the $x^5=0$
$SO(10)$ fixed line.  Under the ${\cal Z}_2$ reflection we assign
parities $H_{\bf 10}(+)$ and $H^c_{\bf 10}(-)$ without loss of
generality.  Under the 4-2-2 gauge symmetry, $H_{\bf 10}$ decomposes
as $({\bf 1},{\bf 2},{\bf 2})+({\bf 6},{\bf 1},{\bf 1})$, 
and under the ${\cal T}_2$ translation, these components have opposite
parity; with the proper choice of sign only the $({\bf 1},{\bf 2},
{\bf 2})$ piece, containing the two Higgs doublets of the MSSM, has a
massless zero mode.  These massless fields couple to the quarks and
leptons through the $SO(10)$ brane superpotential term 
$\psi_{\bf 16} \psi_{\bf 16} H_{\bf 10}$.  Supersymmetry breaking can be 
localized, for instance, to the 3-2-1-1 brane and mediated by the bulk
gauginos.  The $U(1)_X$ breaking can occur on the 5-1 brane and can be
communicated to the $SO(10)$ brane by $\chi_{\bf 16} +
\barr{\chi}_{\barr{\bf 16}}$ pairs propagating in the $x^6=0$ fixed line
as described for the $T^2/Z_2$ model in subsection \ref{subsec:qlb}.  

One property of this model is that $SO(10)$ mass relations hold.  By
mixing the brane-localized $\psi_{\bf 16}$'s with ${\bf 16}$s
propagating on the $x^6=0$ fixed line using the mechanism of 
Ref.~\cite{Hall:2001pg}, these relations can be corrected, but $SU(5)$
mass relations still hold.  This remaining $SU(5)$ mass relations are
corrected by further mixing with states on the $x^5=0$ line.  A
different model with realistic fermion masses is given by starting with
the quarks and leptons contained in $\psi_{\bf 16}$'s propagating on the 
$x^6=0$ fixed line. Depending on its parity under ${\cal T}_1$
translations, each $\psi_{\bf 16}$ contains a zero mode for either 
$\barr{\bf 5}_{3}$ and ${\bf 1}_{-5}$ or for ${\bf 10}_{-1}$, where the
multiplets are labelled by their transformation properties under $SU(5)
\otimes U(1)_X$.  Cancellation of 4D anomalies then requires these 
$\psi_{\bf 16}$'s to appear in pairs with opposite ${\cal T}_1$
parities, so that each pair yields a full generation.  The Higgs
multiplets can propagate on either the $x^5=0$ or $x^5=\pi R_5$ lines.
Taking them to appear in a hypermultiplet ${\bf H}_{\bf 10}$ propagating
on the $x^5=0$ line as before, fermion masses again arise from 
$\psi_{\bf 16} \psi_{\bf 16} H_{\bf 10}$ superpotential terms localized
on the $SO(10)$ fixed point.  This time, the fermion mass relations are
those of $SU(5)$ before mixing with bulk states. These relations can be
corrected by mixing components of the $\psi_{\bf 16}$'s with states
propagating on the the $x^5=\pi R_5$ fixed line (which feel the $SU(5)$
breaking through the ${\cal T}_2$ operation), through couplings on the
5-1 fixed point.  Gaugino mediation can be realized in this model by
localizing the supersymmetry breaking to the 4-2-2 point.  The $U(1)_X$
breaking can again occur on the 5-1 point, and this time the
right-handed neutrinos couple to the breaking directly to pick up their
masses.

\section{Gauge Coupling Unification}
\label{sec:gcu}

In the zero-th order approximation, successful gauge coupling
unification is achieved in these models by identifying the
compactification scale with the unification scale, $1/R \sim M_U = 2
\times 10^{16}$ GeV.  There are, however, two types of corrections to
this naive identification \cite{Hall:2001pg, Nomura:2001mf}. 

First, we can write down tree-level gauge kinetic terms that do not
respect the full $SO(10)$ symmetry on subspaces of the 6D spacetime.  As
an example, we can write 5D gauge kinetic terms respecting only 
$SU(4)_C \otimes SU(2)_L \otimes SU(2)_R$ gauge symmetry on the 
$x^6 = \pi R_6$ fixed line in the $T^2/(Z_2 \times Z_2')$ model of
section \ref{section:z2sq}.  Also, 4D gauge kinetic terms are introduced
on each orbifold fixed point, which need only respect the gauge
symmetries remaining there.  However, the corrections from these
operators are generically suppressed by the volume of the extra
dimension(s), so that we will neglect these contributions in the
following discussion.

The second correction originates from the running of the gauge couplings
above the compactification scale due to KK modes.  Since the present
model is a 6D theory, the zero-mode gauge couplings $g_{0i}$ at the
compactification scale $M_c$ ($\equiv 1/R$) receive power-law
corrections as \cite{Dienes:1998vh}
\begin{eqnarray}
  \frac{1}{g_{0i}^2(M_c)} \simeq 
    \frac{1}{g_0^2(M_*)} 
    - \frac{b}{8 \pi^2} ((M_* R)^2 - 1)
    - \frac{b'_i}{8 \pi^2} (M_* R - 1)
    + \frac{b''_i}{8 \pi^2} \ln(M_* R),
\label{eq:powerlaw}
\end{eqnarray}
where $b, b'_i$ and $b''_i$ are constants of $O(1)$ and $M_*$ is the 
cutoff scale of the theory.  In the 6D picture, the last three terms
correspond to 6D, 5D and 4D gauge kinetic terms generated by loop
effects in the 6D bulk, on the 5D fixed lines and on the 4D fixed
points, respectively.  An interesting fact is that for the models
possessing 6D $N=2$ supersymmetry in the bulk, the term quadratically
sensitive to the cutoff does not appear, $b=0$.  On the other hand, for
the $T^2/Z_2$ model of section \ref{section:z2} the term quadratically
sensitive to the cutoff does appear, but the crucial point is that this
term is universal, and will not affect the differences between the gauge 
couplings. In fact, since the bulk $SO(10)$ gauge symmetry is spoiled
only at 4D fixed points, the differential running of the gauge couplings
above the compactification scale will be logarithmic, and threshold
corrections to $\sin^2\theta_w$ will be small.

The same conclusion can be drawn for the $T^2/Z_6$ model of section
\ref{section:t2z6}, which also has $SO(10)$ gauge symmetry everywhere
but on 4D fixed points.  The story is different, however, for the
$T^2/(Z_2 \times Z_2')$ model, which has fixed lines with reduced gauge
symmetry.  Although $b=0$ is guaranteed for this model due to the 6D
$N=2$ supersymmetry, the $b_i'$ do not vanish, and moreover are not
universal. As a consequence the gauge couplings experience power-law
(linear) differential running above the compactification scale.
Threshold corrections to $\sin^2\theta_w$ become quite large if the
cutoff is taken to be much larger than the compactification scale, and
we estimate this correction to be $\sim (2-3)\%$ for $M_*R \sim 3$.
In the $T^2/(Z_2 \times Z_2')$ model, consistency with low-energy data
most likely requires some degree of cancellation between threshold
corrections coming from unknown cutoff-scale physics and this correction
arising from KK modes.

\section{Conclusions}
\label{sec:concl}

In this paper we constructed three supersymmetric $SO(10)$ theories in
which the gauge symmetry is broken by orbifold compactification.
Unlike in the $SU(5)$ case, where a single extra dimension is sufficient
for breaking the gauge symmetry, we find in the $SO(10)$ case that at
least two extra dimensions are required to break the symmetry to 
$SU(3)_C \otimes SU(2)_L \otimes U(1)_Y \otimes U(1)_X$.
In each of these theories, the orbifold allows an elegant solution 
to doublet-triplet splitting, and removes proton decay from colored 
Higgsino exchange.

Since we are led to consider 6D theories, important obstacles absent in
the 5D case arise. One is the potential for bulk anomalies, present
only in theories with even spacetime dimensions.  A second is the
possibility of subspaces with reduced gauge symmetry that spoil
successful gauge coupling unification.  In 5D theories the subspaces
with reduced gauge symmetry are 4D, giving only a logarithmic threshold
correction to $\sin^2\theta_w$.  Depending on the orbifold, 6D theories
may have 5D subspaces on which the gauge symmetry is broken, leading to
a power-law correction to $\sin^2\theta_w$.  A different challenge,
particular to $SO(10)$ theories, is that the orbifold compactification
generically does not break the gauge symmetry all the way down to the
standard model gauge group, as it can in 5D $SU(5)$ theories: $U(1)_X$
is left unbroken, and the right-handed neutrinos are massless.

The first model is constructed on a $T^2/Z_2$ orbifold, and possesses 6D
$N=1$ supersymmetry.  The structure of the orbifold is such that the
full $SO(10)$ is realized everywhere but on 4D fixed points,
guaranteeing that threshold corrections to $\sin^2\theta_w$ are under
control. The irreducible bulk gauge anomalies can be canceled by adding
two bulk hypermultiplets in the fundamental representation (containing
the two Higgs doublets of the MSSM), with the option of adding additional 
pairs of hypermultiplets, each pair containing one spinor and one
fundamental.  This allows one to build models in which doublet-triplet
splitting is naturally realized, and in which $U(1)_X$ is broken by the
VEV of an $SU(5)$ singlet localized on an $SU(5) \otimes U(1)_X$
preserving brane, giving rise to masses for the right-handed neutrinos
either through direct interaction (for the case of matter in the bulk),
or by integrating out bulk states (for the case of matter on a fixed
point).  Even for the case of matter localized to an $SO(10)$ preserving
fixed point, unwanted GUT fermion mass relations can be corrected
through mixing with bulk states.  Anomaly cancellation requires the 
Green-Schwarz mechanism \cite{Green:1984sg}, which leads to light states 
with axion-like couplings.  This model is fully realistic.

We also considered theories with 6D $N=2$ supersymmetry, 
for which complete anomaly cancellation is automatic.  The
$T^2/Z_2$ orbifold cannot be used to build such a theory because the
fixed points have too much supersymmetry left over after orbifolding: 4D
$N=2$ rather than 4D $N=1$.  Instead we used a $T^2/Z_6$ orbifold that
gives two points with 4D $N=1$ supersymmetry, and $SU(4)_C \otimes
SU(2)_L \otimes SU(2)_R$ and $SU(5) \otimes U(1)_X$ gauge symmetries,
respectively.  As in the $T^2/Z_2$ model, the $SO(10)$ is broken only
at 4D points and so there is no power-law differential running of the
gauge couplings above the compactification scale.  To avoid light
colored Higgs states the most convenient choice is to put matter and
Higgs on the 4-2-2 point.  The $U(1)_X$ symmetry can be broken by the
VEV of an $SU(5)$ singlet on the 5-1 brane, but there is no clear way of
communicating this breaking to the 4-2-2 brane to give right-handed
neutrino masses; also, there is no clear way of relaxing $SO(10)$
fermion mass relations.  The alternative of breaking $U(1)_X$ on the
4-2-2 fixed point is also problematic as it leads to a vacuum alignment
problem and massless Goldstone states: canceling anomalies by
restricting the bulk matter content to be 6D $N=2$ supersymmetric makes
$U(1)_X$ communication and attainment of realistic fermion masses a
challenge because it makes the bulk less accessible.

We explored a resolution to this problem in the third model, on 
$T^2/(Z_2 \times Z_2')$.  Although the bulk is again taken to possess 6D 
$N=2$ supersymmetry, this orbifold has 5D lines, fixed under one $Z_2$
but not the other, which possess only 6D $N=1$ supersymmetry.  These
lines connect 4D points with 4D $N=1$ supersymmetry, where interactions
can arise.  Matter multiplets may be introduced on these 5D lines
without spoiling bulk anomaly cancellation, so this set up yields a
number of possibilities for realistic models.  In particular, natural
doublet-triplet splitting is accommodated by appropriate placement of
Higgs multiplets on these lines, communication of $U(1)_X$ breaking is
now straightforward, and realistic fermion masses can be attained.  
The trade-off is that this model has 5D surfaces that do not preserve
$SU(5)$, leading to linear running of the gauge couplings relative to
one another above the compactification scale: the fixed lines are
welcome for certain model building purposes but damaging for the
prediction of $\sin^2\theta_w$, especially if the cutoff is taken much
larger than the compactification scale.  An attractive $N=2$ model would
be one with only $SO(10)$ or $SU(5) \otimes U(1)_X$ preserving fixed
lines.  The challenge is to realize this situation while breaking the
gauge symmetry down to the standard model group, and while accommodating
natural doublet-triplet splitting, right-handed neutrino mass
generation, and realistic fermion mass matrices.  It will be interesting
to pursue a fully realistic model along these lines in the future.

\end{document}